\documentclass[11pt]{amsart}

\usepackage{amsmath,amssymb,amsthm,mathtools}
\usepackage[colorlinks=true,linkcolor=blue,citecolor=blue,urlcolor=blue]{hyperref}
\usepackage[round]{natbib}

\theoremstyle{plain}

\theoremstyle{definition}

\theoremstyle{remark}

\numberwithin{equation}{section}

\newcommand{\R}{\mathbb R}

\title[By Law, Every Zero-Mean Risk Is the Difference of Two Equally Distributed Risks]{By Law, Every Zero-Mean Risk Is the Difference of Two Equally Distributed Risks}
\author{Mark Whitmeyer}
\date{\today}
\thanks{\emph{Acknowledgements.} Dedicated to KS. I used ChatGPT for proof-checking.}

\begin{document}

\begin{abstract}
    We prove that every mean-zero law on \(\R\) is the law of \(X-Y\) for some identically distributed real-valued random variables \(X\) and \(Y\).
\end{abstract}

\maketitle

Following the setup of \citet{maccheroniinsurancerisk}, and letting \(\mu=P_f\) with \(\int z \mu(dz)=0\); for nonatomic \(P\), set \(p_0=\mu(\{0\})\), \(\mu_+=\mu|_{(0,\infty)}\), \(\mu_-=(-\mathrm{id})_\#(\mu|_{(-\infty,0)})\), and \(a=\int_0^\infty x \mu_+(dx)=\int_0^\infty y\,\mu_-(dy)\); then take \(\nu=\delta_{(0,0)}\) if \(a=0\), otherwise, let \(R_{x,y}\) be rotation by \(x\) on the circle \([-y,x)\) and define
\[
\nu
=
p_0\delta_{(0,0)}
+
\frac1a
\int_{(0,\infty)^2}
\int_{-y}^{x}
\delta_{(R_{x,y}u,u)}\,du\,\mu_+(dx)\mu_-(dy),
\]
so, by nonatomicity, choose \(X,Y\) on \(S\) with \(P_{(X,Y)}=\nu\); then \(X\stackrel d=Y\) and
\[\begin{split}
    P_{X-Y} &= p_0\delta_0 + \frac1a \int_{(0,\infty)^2} \left(y\delta_x+x\delta_{-y}\right)\,\mu_+(dx)\mu_-(dy)\\
    &= p_0\delta_0+\mu_+ + (-\mathrm{id})_\#\mu_- = \mu = P_f.
\end{split}\]

This proves the nonatomic case of Lemma 1 in \citet{maccheroniinsurancerisk}.\footnote{Mostly; it remains to check that \(X,Y\in\mathcal F\). If \(a=0\), then \(\nu=\delta_{(0,0)}\), so \(X=Y=0\) a.s. Assume \(a > 0\). If \(f\in \mathcal{L}^\infty\), choose \(M<\infty\) with \(\mu([-M,M])=1\), so that \(\mu_{+}((M,\infty))=\mu_{-}((M,\infty))=0\). Accordingly, \(\nu([-M,M]^2)=1\); indeed, for \(0<x,y\le M\) and \(u\in[-y,x)\), \(u,R_{x,y}u\in[-M,M]\). Since \(P_{(X,Y)}=\nu\), \(X,Y\in \mathcal L^\infty\). If \(f\in \mathcal{M}^\infty\), then for every \(r\ge1\),
\[\mathbb E[|Y|^r]
=
\frac{1}{a(r+1)}
\left[
\mu_-((0,\infty))\int_0^\infty x^{r+1}\mu_+(dx)
+
\mu_+((0,\infty))\int_0^\infty y^{r+1}\mu_-(dy)
\right]<\infty,
\]
and the same holds for \(X\) since \(X\stackrel d=Y\), so \(X,Y\in \mathcal{M}^\infty\). \hfill \(\blacksquare\)} The finite-uniform case is the cyclic partial-sum construction they provide; and these two cases exhaust their maintained adequacy assumption on \(P\).

\bibliographystyle{plainnat}
\bibliography{references}

\end{document}